# Transitioning from file-based HPC workflows to streaming data pipelines with openPMD and ADIOS2

Franz Poeschel[1,4], Juncheng E[5], William F. Godoy[3], Norbert Podhorszki[3], Scott Klasky[3], Greg Eisenhauer[6], Philip E. Davis[7], Lipeng Wan[3], Ana Gainaru[3], Junmin Gu[2], Fabian Koller[4], René Widera[4], Michael Bussmann[1,4], and Axel Huebl[2,4]

[1] Center for Advanced Systems Understanding (CASUS), D-02826 Görlitz, Germany
[2] Lawrence Berkeley National Laboratory (LBNL), Berkeley 94720, California, USA
[3] Oak Ridge National Laboratory (ORNL), Oak Ridge 37830, Tennessee, USA
[4] Helmholtz-Zentrum Dresden-Rossendorf (HZDR), D-01328 Dresden, Germany
[5] European XFEL GmbH (EU XFEL), D-22869 Schenefeld, Germany
[6] Georgia Institute of Technology (Georgia Tech), Atlanta 30332, Georgia, USA
[7] Rutgers University (Rutgers), New Brunswick 08901, New Jersey, USA

**Abstract.** This paper aims to create a transition path from file-based IO to streaming-based workflows for scientific applications in an HPC environment. By using the openPMP-api, traditional workflows limited by filesystem bottlenecks can be overcome and flexibly extended for in situ analysis. The openPMD-api is a library for the description of scientific data according to the Open Standard for Particle-Mesh Data (openPMD). Its approach towards recent challenges posed by hardware heterogeneity lies in the decoupling of data description in domain sciences, such as plasma physics simulations, from concrete implementations in hardware and IO. The streaming backend is provided by the ADIOS2 framework, developed at Oak Ridge National Laboratory. This paper surveys two openPMD-based loosely-coupled setups to demonstrate flexible applicability and to evaluate performance. In loose coupling, as opposed to tight coupling, two (or more) applications are executed separately, e.g. in individual MPI contexts, yet cooperate by exchanging data. This way, a streaming-based workflow allows for standalone codes instead of tightly-coupled plugins, using a unified streaming-aware API and leveraging high-speed communication infrastructure available in modern compute clusters for massive data exchange. We determine new challenges in resource allocation and in the need of strategies for a flexible data distribution, demonstrating their influence on efficiency and scaling on the Summit compute system. The presented setups show the potential for a more flexible use of compute resources brought by streaming IO as well as the ability to increase throughput by avoiding filesystem bottlenecks.

## 1 The need for loosely-coupled data pipelines

Scientists working with massively scalable simulations on high-performance compute (HPC) systems can currently observe an increasing IO bottleneck threatening the performance of their workflows. As GPU hardware has reshaped the compute landscape found in the TOP500 list[8], state-of the art HPC codes have become able to exploit the compute power of thousands of GPUs in parallel. When storage systems cannot keep pace with this development, workflows must be adapted to continue exploiting advancements made in compute performance. This paper explores streaming IO as a scalable alternative to persistent IO.

This section first gives an overview on the performance of state-of-the-art supercomputers as well as on typical scientific workflows for massive data processing. Section 2 proposes streaming IO as an approach at keeping these workflows scalable on recent systems. Section 3 discusses the challenge of streaming data distribution. Finally, section 4 builds and examines two prototypical streaming data processing pipelines and evaluates the data distribution patterns previously discussed.

### 1.1 The IO bottleneck – a challenge for large-scale IO

Hoping to bring forward more detailed scientific insights, recent supercomputer systems strive to enable simulation sizes that are impossible to fit on smaller clusters. Applications that use a large percentage of resources on such a system are challenged to near-perfect parallel weak scaling. While from the perspective of the compute system, achieving this goal – while demanding – has been proven possible by applications such as PIConGPU [3], storage systems on recent systems are far less scalable:

| system | compute performance [$\text{PFlop} \cdot \text{s}^{-1}$] | parallel FS bandwidth [$\text{TiByte} \cdot \text{s}^{-1}$] | FS capacity [PiByte] | example storage requirements [PiByte] |
|---|---|---|---|---|
| **Titan** | 27 | 1 | 27 | 5.3 |
| **Summit** | 200 | 2.5 | 250 | 21.1 |
| **Frontier** | > 1500 | 5 - 10 | 500 - 1000 | 80 - 100 |

Table 1: System performance: OLCF Titan to Frontier. The last column shows the storage size needed by a full-scale simulation that dumps all GPU memory in the system 50 times.

Table 1 shows that the full-scale peak compute performance increases by a factor of ∼ 7.4 from Titan (2013) to Summit (2018) and by a further factor of > 7.5 from Summit to Frontier (planned 2021). Conversely, the parallel bandwidth increases from Titan to Summit by merely 2.5 and the parallel bandwidth of

---

[8] https://www.top500.org/

Frontier is planned at 2-4 times that figure. For the storage capacity, the increase from Titan to Summit goes by a factor of 7.8, keeping up with the increase in compute performance. The storage capacity for Frontier, however, is planned at 2-4 times that of Summit, falling behind the pace of the peak compute performance. The table shows that full-scale applications that write regular data dumps will use significant portions of the system storage.

Large-scale capability application runs perceive this as an *IO wall*: At full scale, the theoretical maximum parallel filesystem (PFS) throughput per node on Titan (NVidia Tesla K20x) is $56\,\text{MByte} \cdot \text{s}^{-1}$, and $95\,\text{MByte} \cdot \text{s}^{-1}$ on Summit (NVidia Tesla V100). Summit compute nodes are further equipped with 1.6 TiB of non-volatile memory (NVM). While envisioned as a burst-buffer for the PFS, draining of NVM data dumps during parallel application runs can lead to a competition for network resources.

Traditional methods for further processing of simulation data reach their limits due to the PFS bandwidth. Simulations intending to scale up to full capability cannot continue to rely on file-based IO and must explore alternative methods for further processing of data.

## 1.2 From monolithic frameworks to loosely-coupled pipelines

Multi PB-scale data output does not solely pose a technical challenge, as from a domain scientist's perspective this raw data also provides a relatively low informational density, necessitating data extraction through adequate analysis.

An example is the particle-in-cell simulation PIConGPU [3, 9], developed at the Helmholtz-Zentrum Dresden-Rossendorf. It produces massive amounts of raw data that requires further processing and extraction of information. As depicted in figure 1, PIConGPU provides a number of tightly-coupled *plugins* to perform simple analysis tasks on raw data.

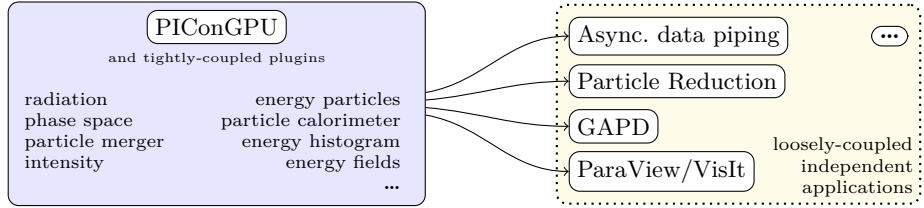

Fig. 1: Tight and loose coupling in the example of PIConGPU

In the context of this work, the *tightly-coupled* approach is defined as a data analysis routine built into the data producer, which cannot be used as a standalone application. It brings the advantage of trivial data access at the expense of flexibility. While a tightly-coupled plugin can benefit from using the same address space as the actual simulation, it also comes with a list of downsides. The development of such plugins is highly application specific, requiring more development resources, and the software stack of all plugins, including potentially very large frameworks with incompatible dependencies and compiler contraints, needs to compile and link at once. The heavy investment of developing a plugin only pays off for algorithms that scale well with the simulation size. As shown in [14], algorithms that require communication between the processes can sometimes run faster separately at smaller scale even with the additional cost of data movement. An example analysis in large 3D PIConGPU runs that does not benefit from proportionally scaled-up parallel analysis is the extraction of a distributed 2D simulation slice followed by a 2D Hilbert transform, often used to find the position of a laser pulse.

This paper defines *loose coupling* as a workflow wherein multiple discrete stages in a computational data processing pipeline are performed by different independent applications, usually launched as different processes on the system. Loose coupling does not require combining several heterogeneous stages of a simulation/analysis/visualization/aggregation workflow into a monolithic integrated application, which is indeed often impossible due to contradicting dependency requirements in complex software stacks.

This paper's approach at creating loosely-coupled workflows in an efficient and scalable manner is data streaming, introduced in section 2.

### 1.3  Related work

SIMEX platform [6, 7] implements the openPMD standard [11] through HDF5 files and uses loose coupling to integrate "some of most advanced simulation tools [...] to mimic an entire light source beamline". Similarly, Mayes et al. [19] (LUME) integrates single simulation modules with openPMD into a file-based loosely-coupled pipeline for unified particle accelerator and lightsource modeling. Wan et al. [22] investigate layout reorganization of parallel PIC simulation data to increase file IO performance, including downstream processing and data loads on finely chunked, distributed data. An established framework for loose coupling in the domain of Big Data is the JVM-based Apache Spark [24]. DASK [4] is a Python-based framework for "scalable analytics". Its approach is to couple multiple loosely-coupled tasks via the definition of a fine-grained task-graph, yielding an acyclic dataflow. FLUX [2] builds a framework for the graph-based scheduling for jobs in an HPC environment, allowing to divide compute resources across multiple applications. The position of this paper's work in frameworks such as DASK or FLUX is that of a means for exchange of scientific data between arbitrarily scaled, parallel compute/analysis processes controlled by these emerging frameworks. Staging (streaming) methods in ADIOS are described in Abbasi et al. [1] and were recently redesigned into a scalable publish/subscribe abstraction that supports various coupling use-cases [17]. Both Ascent (part of the Alpine project [15]) as well as the SENSEI [18] project build in-situ visualization and analysis algorithms, running alongside simulations to avoid bottlenecks in bandwidth and storage incurred by post-hoc processing. ADIOS is an available data transport method. Compared to these frameworks, openPMD initially standardizes data organization and self-describing meta-data as frequently needed for analysis, coupling and interpretation in the respective domain-science [11]; it can be implemented directly on a variety of data transports/libraries and formats, yet also provides a scalable reference implementation, openPMD-api [13].

## 2 Building a system for streaming IO

The considerations in subsection 1.2 already hint at the largest challenge in loose coupling compared to tight coupling: Instead of reducing the raw simulation data by mode of analysis within the simulation by using a tightly-coupled plugin, transfer of massive amounts of data into a reading application becomes necessary, rendering relevant the IO bottleneck discussed in subsection 1.1. This section presents *data streaming* (sometimes also referred to as *staging*) as a means to implement highly-scalable loosely-coupled data processing pipelines.

### 2.1 Loosely coupled data processing pipelines built via streaming

*Data streaming* in this work refers to a form of IO between two (or more) applications that bypasses persistent storage media, using instead scalable high-speed communication infrastructure such as Infiniband. In such workflows, data is sometimes held (staged) in a temporary location, e.g. RAM or NVM.

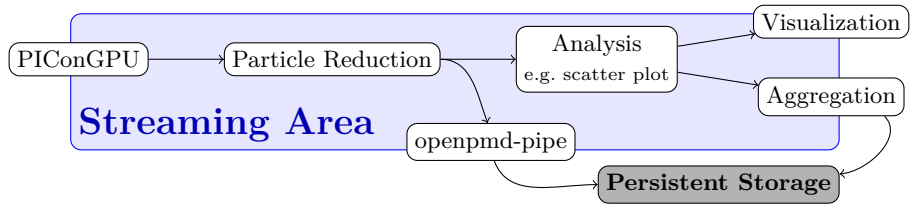

Fig. 2: An envisioned complex, loosely-coupled workflow: PIConGPU is the data producer, a domain-specific particle reduction can conserve relevant ensemble properties, the analysis step might filter and bin, and aggregation might create a temporal integration from high-frequency data. At various sections of the workflow, visualization or data dumps might be generated from subscribers.

Such methods are interesting from two perspectives: For the scientist working in a loosely-coupled setup, persistent storage of intermediate raw data is in general not necessary, and from the perspective of the IO system, it is not feasible. In consequence, our response to the IO wall lies in streaming-based loosely-coupled workflows that allow to forgo persistent storage infrastructure bound by parallel bandwidths and filesystem capacities. An example for a loosely-coupled scientific data processing pipeline, consisting of multiple simulations, is depicted in figure 2, demonstrating a multi-stage pipeline that avoids writing unwanted intermediate results to permanent storage. The script `openpmd-pipe`, which allows for capturing a stream into a file, serves as basis for this paper's first benchmark in section 4.1.

The IO system presented in this work focuses on the following criteria, driven by needs of domain scientists:

**Efficiency** Naturally, any IO system prepared for the Exascale era needs to move massive amounts of data in a scalably efficient manner.

**Expressiveness** Scientists need to express, transport and share their data and meta-data in a language that comes naturally within the problem's domain. Any need for manually dealing with low-level concepts should be avoided and needs of FAIR principles [23] should be considered as integral part.

**Flexibility** The IO system should be usable on different systems and in different setups with different IO backends and configurations without altering the application code's data description. Optimization for system specifics should be exposed through runtime configuration.

**Reusability** Current-day IO routines are often written in terms of persistent-storage based reading and writing. Upgrading to a streaming-aware IO description should be straightforward. The upgraded routine should allow for file-based as well for streaming-based IO.

## 2.2 Impact in an increasingly heterogeneous compute landscape

These properties in an IO system prepare scientific workflows for a trend towards heterogeneity in software and hardware, up to approaches such as federated computing. Keeping true to the efficiency of integrated solutions, this more modular approach avoids a fixed description of the concrete IO layer, replacing it with an idiomatic data-focused description within the scientific domain and leaving the physical choice of IO transport or storage as a runtime parameter. While the remainder of this paper focuses on two prototypical setups for performance and feasibility benchmarking, further more complex workflows are thinkable, two examples including edge computing and machine learning workflows.

For edge computing, such an IO framework, while still satisfying the needs of classical scientific simulation workflows, can serve as the communication layer in the complex communication patterns demanded between compute and edge nodes. Scientific software is thus generalized to unify traditional as well as upcoming workflows into one data description. Similarly, such an IO system makes it possible to schedule each compute part of a loosely-coupled simulation separately on heterogeneous systems, making use of available compute resources in a way best fit for each one. Hence, a GPU compute partition can be used for hardware-accelerated simulation codes, while a CPU-based post-processing transformation of the data can be scheduled on more fitting hardware.

An interesting application for machine learning is found in the computation of surrogate models for simulations. As argued before, scientific simulations create lots of data – in machine learning setups, the training of an accurate model additionally needs a rich set of input data. Both these facts combine into data requirements that can no longer be supported by file-based IO routines. An envisioned workflow supported by a flexible IO system consists of a long-running surrogate model training process being fed by dynamically launching instances of a scientific simulation via streaming methods, thus bypassing the filesystem.

### 2.3 openPMD and ADIOS2: Scientific Self-Description & Streaming

In this paper, we explore a streaming-aware IO infrastructure that leverages scientific self-description of data through the Open Standard for Particle-Mesh Data (openPMD) [11]. We use ADIOS2 [8] to add a streaming backend implementation to the library openPMD-api [13].

The openPMD-api, developed openly in collaboration of Helmholtz-Zentrum Dresden-Rossendorf, the Center for Advanced Systems Understanding and Lawrence Berkeley National Laboratory, is an IO middleware library that assists domain-scientists with data description along the openPMD standard for FAIR particle-mesh data, used already in numerous physics simulations.[9]

In comparison to direct implementations of application codes against high-speed IO backends [10], using openPMD-api saves thousands of lines of code per application, reduces integration time for application developers, promotes sharing of best practices for IO library tuning options, provides a high-level interface to describe scientific data and conserves this standardized meta-data for analysis and coupling workflows. The *expressiveness* property from subsection 2.1 is thereby achieved. The ability to pick different backends and configurations for them at runtime brings *flexibility*. Implementing high-speed backends such as HDF5, ADIOS1 and ADIOS2 as depicted in figure 3 achieves *efficiency* (whereas a serial JSON backend serves for prototyping and learning purposes). Additional language bindings on top of the core C++ implementation ease integration into many established post-processing frameworks, such as Python bindings for parallel readers into ParaView, Dask and domain-specific analysis tools.

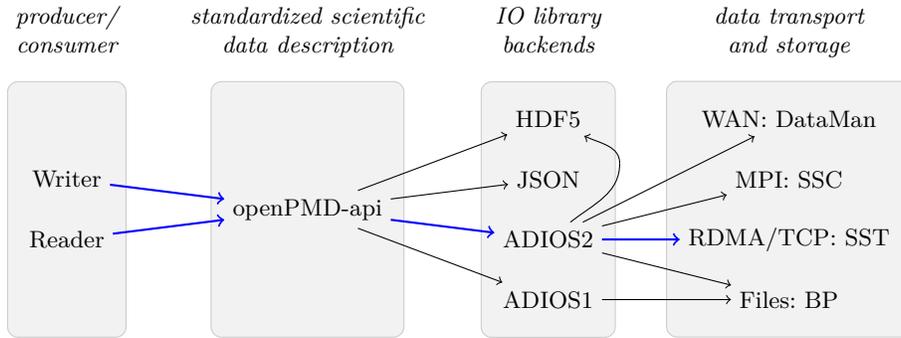

Fig. 3: IO software stack built by openPMD and ADIOS2

We leverage the benefits of the openPMD-api and make its interface aware of streaming, thus allowing scientists that already script their traditional data analysis routines for file-based IO to rapidly transition to streaming IO with a small number of changes, keeping in mind the goal of *reusability*.

The ADIOS2 (*adaptable IO system*) framework [8], developed by Oak Ridge National Laboratory, in collaboration with Kitware Inc., Lawrence Berkeley

---

[9] https://github.com/openPMD/openPMD-projects

National Laboratory, Georgia Institute of Technology and Rutgers University, is a unified high-performance IO framework, located one logical layer of abstraction below the openPMD-api, and provides the backend implementation for streaming IO. It supersedes the earlier ADIOS1 framework [16] by creating a general IO-API based on the publisher/subscriber model, decoupling the data producers and consumers in a manner that allows for a flexible framework to incorporate various data movement *engines* specialized for different use cases (permanent storage, staging through local NVMes, streaming through RDMA, TCP and MPI, etc). It provides *efficiency* at extreme-scale and *flexibility* by providing many specialized IO engines. ADIOS2 also supports *reusability* via the publish/subscribe API that allows for selecting an engine at runtime without any change to the user code.

The openPMD-api (0.14.0) uses mainly its BP3 and BP4 (*binary pack*) engines for file-based IO, SSC (*strong static coupler*) for strong coupling via MPI, and its SST (*sustainable staging transport*) engine for loose coupling [17] on which this paper focuses.

The purpose of SST is to enable very dynamic coupling workflows, allowing arbitrary numbers of readers to register to a stream while it keeps running. Between each writer and reader, communication between their respective parallel instances can go in arbitrary patterns up to full $m \times n$ meshes, opening connections only between instances that exchange data. The engine can pick from different low-level data transport implementations: The libfabric[10]-based RDMA (*remote direct memory access*) data transport for high scalability and use of technologies such as Infiniband, as well as the TCP data transport that uses network sockets for communication and that works as a fallback if libfabric is not available on the system.
The openPMD-api documentation[11] lists and explains a minimal example how to communicate between a writer and a reader with ADIOS2/SST (highlighted blue in figure 3). If not specified differently, this study uses the SST engine with the RDMA data transport, backed by libfabric 1.6.3a1.

## 3  Data distribution patterns

A writer for openPMD-formatted output will generally produce data in form of *n*-dimensional chunks that differ in size (location in the problem domain) and parallel instance of origin (location in the compute domain). A non-trivial decision problem emerges concerning which data regions should be read by which instance of a reading code. This section determines properties that efficient data distributions should have and discusses a number of algorithms to achieve these properties.

---

[10] https://ofiwg.github.io/libfabric/
[11] https://openpmd-api.readthedocs.io/en/0.13.3/usage/streaming.html

### 3.1 Properties found in a performant distribution pattern

The SST engine of the ADIOS2 framework leaves some leeway for experimentation and optimization by theoretically allowing fully-connected communication patterns where each reading instances holds communication with each writing instance. For a performant pattern, we put forward a number of properties:

**Locality** Rather than fully-connected "$m \times n$" style communication patterns, each instance's communication partners should bounded in number and located close within the system's topology. When carefully picking a local communication pattern, perfect IO weak scaling behaviour can ideally be achieved. Naturally, this point is subject to the communication infrastructure and the topology of a system.

**Balancing** The distribution of data over parallel instances of a reading application should be as even as possible. Loose coupling serves as an opportunity to implement load balancing in a data processing pipeline.

**Alignment** Access patterns should be congruent with the data layout in the backend. For ADIOS2, which generally organizes data in form of the chunks as they were written, this means that loaded chunks should coincide as much as possible with those that were written. Loading a chunk that must be pieced together from parts of several written chunks inhibits efficiency.

**Read constraints** Applications often impose domain-specific constraints on the temporal and/or spatial order of data for processing. This can limit the sensible distribution patterns.

### 3.2 Chunk distribution algorithms

This subsection surveys a number of chunk distribution algorithms in the context of the properties as detailed in subsection 3.1. Pseudocode is found in the supplementary material [20]. Since the *read constraints* property depends on the data requirements within the problem domain, we do not further concern ourselves with it in this subsection.

Each of the following algorithms guarantees a complete distribution of data from writers to readers. Efficiency cannot generally be guaranteed and requires careful scheduling of applications and selection of a fitting distribution strategy.

**Round Robin** The Round Robin approach at chunk distribution distributes the available data chunks over the parallel readers until none are left. It optimizes only for the *alignment* property, fully forgoing the properties of *locality* and *balancing*. It is interesting only in situations where its effects can be fully controlled by other means, such as knowledge on the way data is produced and consumed.

**Slicing the dataset into n-dimensional hyperslabs** Since datasets have known sizes, one possibility is pre-assigning regions in form of hyperslabs in a dataset to reading ranks. This comes close to the conventional approach of explicitly selecting chunks to load. The available chunks are intersected with the hyperslab assigned to a rank to determine that rank's chunks to load.

This approach optimizes for the *balancing* property, mostly ignoring the other two properties. However, the *locality* property can be achieved if the distribution of parallel compute resources correlates with the problem domain, a condition often met in codes that do not use complex load balancing schemes. Similarly, the *alignment* property is achieved to some extent by the controlled number of cuts on the problem domain.

Since the achieved distribution is generally one that can be easily dealt with in reading analysis codes, this makes this rather simple approach interesting.

**Binpacking** This approach tries to combine the advantages of the first two approaches. The rough proceeding is to calculate an ideal amount of data per rank, slice the incoming chunks such that this size is not exceeded and then distribute those size-fitted chunks across the reading ranks. The last step is non-trivial and an instance of the NP-complete bin-packing problem. Johnson [12] discusses algorithms to approximate the problem. For this paper, we adapt the Next-Fit algorithm to approximate the problem within a factor of 2, i.e. at most the double number of bins is used. For the purpose of chunk distribution, it is simple to modify such that each reading rank gets assigned at worst double the ideal amount. Later benchmarks in subsection 4.3 show that this worst-case behavior, while uncommon, does in practice occur.

Other than Round Robin, this algorithm has a guarantee on data *balancing*, and other than the approach that slices the dataset into hyperslabs, it guarantees that the incoming chunks are not arbitrarily subdivided, but instead at most sliced into fixed-sized subchunks, creating some notion of *alignment*. Both guarantees are weakened compared to those given by the earlier algorithms.

**Distribution by Hostname** This algorithm serves to enrich the previous algorithms by some notion of data *locality*, making use of the information on topology of the writing and reading codes. Its schematics are sketched in figure 4.

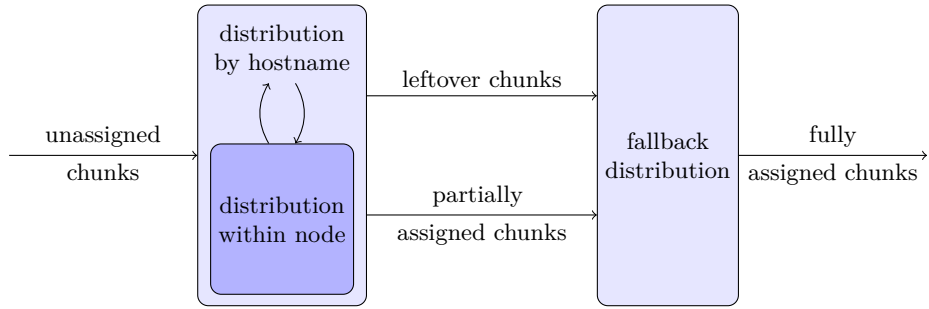

Fig. 4: Workflow for keeping a chunk distribution's *locality* property by keeping communication within the same node.

It works in two phases: The first phase sorts chunks by node to readers on the same node. Specifically, the hostname is used for this step, and it can also be replaced with other layers in a system's topology, such as CPU sockets or host cohorts. A secondary distribution algorithm computes the chunk distribution within each single node.

After this, chunks may remain unassigned if a node is populated only by writers and no readers. To ensure a complete distribution, any of the preceding algorithms can be chosen as a fallback distribution for the leftover chunks.

As a result, this algorithm can be used to dynamically adapt to job scheduling: If nodes are populated by writers and readers at the same time, communication stays within a node. If however a job is scheduled to populate nodes only with either writers or readers, another strategy is automatically picked up.

## 4 Evaluation for two streaming setups

This section sees the construction and evaluation of two setups for streaming. The first setup uses streaming as a method for asynchronous IO, giving us a first insight into streaming performance characteristics. The second setup then constructs a data processing pipeline built from PIConGPU as data source and from GAPD, a GPU-accelerated diffraction simulation, as a data sink. It includes a study on the influence of chunk distribution strategies, of resource allocation, of streaming backend and of parallel scaling.

### 4.1 Streaming as basis for an asynchronous IO workflow

The first setup demonstrates the potential that streaming has for exhausting idle compute resources. Data is streamed to a different process to be written to the filesystem in the background. The approach does not avoid the filesystem bottleneck, but instead provides asynchronous IO and hides the cost of IO if there is enough free memory to stage data.

In our evaluations, the data producer is the aforementioned particle-in-cell simulation code PIConGPU [3]. With the particle-in-cell algorithm being a computational method within the field of plasma physics, the raw output of PIConGPU is particle-mesh based and naturally expressible in openPMD.

For rapid setups of asynchronous IO pipelines via streaming, we developed the tool `openpmd-pipe`, which is an openPMD-api based Python script that redirects any openPMD data from source to sink. While this script performs the most simple transformation that any stage in a loosely-coupled pipeline might possibly do (none at all), it serves as an adaptor within a loosely-coupled pipeline: Further enabled workflows include (de)compressing a dataset or conversion between different backends, and capture of a stream into a file, used by the following setup. Within the context of openPMD, this builds expressivity comparable to using POSIX tee (by using the SST engine as a multiplexer) and to POSIX pipes (by using `openpmd-pipe` as an adaptor piece), motivating its name.

Figure 5 on the next page shows the setup for this first benchmark with `openpmd-pipe` on the Summit supercomputer [21]. (Software versions are documented in the supplementary materials [20].)

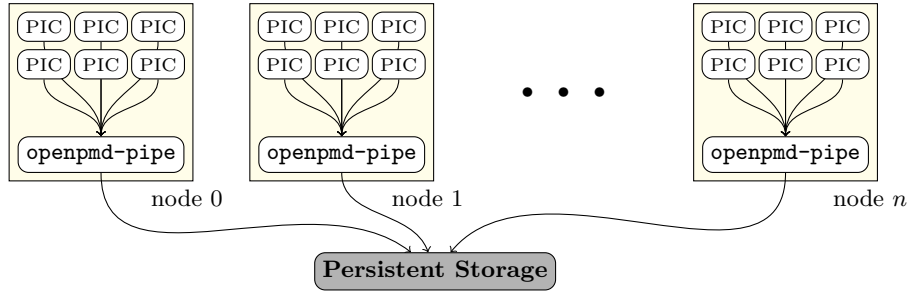

Fig. 5: Benchmark setup: Each node hosts 6 instances of PIConGPU, one per GPU. Those feed their data to one instance of `openpmd-pipe` which then asynchronously writes to the PFS.

Each compute node hosts six instances of PIConGPU, running a Kelvin Helmholtz simulation that is weakly scaled along the $y$ axis. 9.14 GiB are produced per data output step and parallel process. Each node hosts one instance of `openpmd-pipe` to which each PIConGPU process on the same node sends data via SST. `openpmd-pipe` then writes that data to the parallel filesystem (Alpine).

This way, each node creates only one file on the parallel filesystem – a feature also supported natively by the ADIOS2 BP engine under the name of *aggregation*. Regular ADIOS2 BP output with node-level aggregation serves as the benchmarks's baseline. The baseline is referred to as "BP-only", while the setup shown in figure 5 is "SST+BP". Each benchmark runs for fifteen minutes and attempts to write data every 100 time steps in the simulation. The setup uses a feature in the ADIOS2 SST engine to automatically discard a step if the reader is not ready for reading yet.[12]. This way, the simulation is never affected by actual file operations and IO granularity is automatically reduced if it becomes too slow.

Figure 6 on the facing page plots the throughput results. It shows the *perceived throughput* which we define through dividing the amount of data to be stored/sent by the time from starting the operation to its completion. Unlike the raw throughput, this includes latency time needed for communication and synchronization. It provides a good lower bound on the throughput while simplifying measurement in user code. When streaming, the time is measured between requesting and receiving data on the reader's side. Since the SST+BP setup performs IO in two phases, the throughput is shown for both of them.

The throughput is computed by average over each single data dump and over each parallel instance, scaled to the total amount of written data. Each benchmark is repeated three times and the plot shows each of those measurements.

The figure shows reasonable scaling behavior for both filesystem throughput as well as streaming throughput, with slight deviations from ideal scaling appearing at 512 nodes. At the time of running the benchmarks, the streaming example was not able to run without crashing on 1024 nodes.[13] At 4.0 to 4.3 TiByte·s$^{-1}$, the streaming throughput exceeds the PFS bandwidth (2.5 TiByte·s$^{-1}$) at 512 nodes. Since the first phase in the SST+BP setup already performs node-level data aggregation, the perceived BP throughput is higher than in the BP-only setup. At 512 nodes, the PFS bandwidth is approached (2.1 to 2.4 TiByte·s$^{-1}$).

---

[12] `"QueueFullPolicy" = "Discard"` The alternative is to block. A queue of steps can be held for some additional leeway, but it requires additional memory.

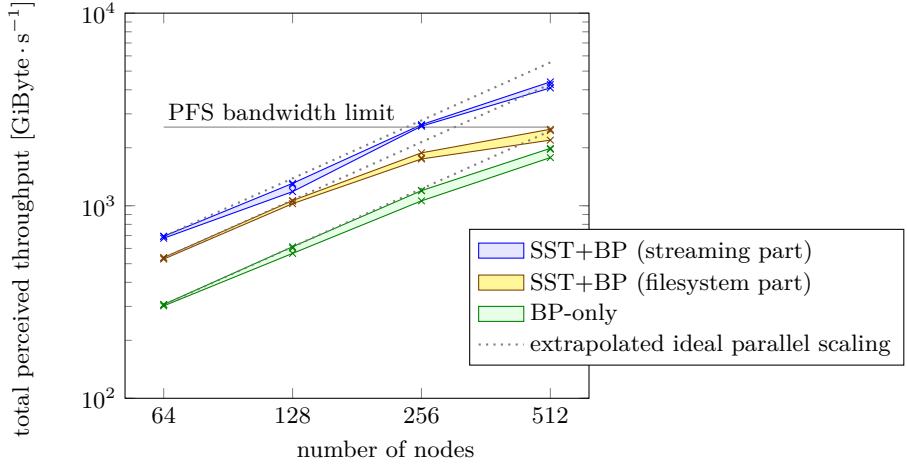

Fig. 6: Perceived total throughput. The file-based outputs (BP-only as well as SST+BP) are limited by the PFS bandwidth. At 512 nodes, the methods reach 4.15, 2.32, and 1.86 TiByte·s$^{-1}$ on average, respectively.

Figure 7 on the next page plots the measured load/store times from the same benchmark. By computing the average numbers, the previous figure did not show any meaningful information on outliers. In parallel contexts, outliers can be serious hurdles for performance, hence figure 7 shows the data as boxplots (including each single measurement across all three repetitions of each benchmark). Results are shown for the streaming part of the SST+BP setup and for the BP-only setup. For BP-only, the median times range between 10 and 15 seconds with the worst outlier at 45 s. The median time for streaming is between 5 and 7 seconds with the worst outlier just above 9 s. A general trend is the increasing number of outliers at 256 nodes. At 512 nodes, longer load times become numerous enough to skew the median and no longer be outliers, explaining the observed decreasing throughput.

Finally, to compare overall performance between both approaches, we count the number of successfully written data dumps within the given time frame of 15 minutes. The BP-only setup blocks the simulation during IO. The number of successfully dumped simulation steps ranges from 22 ∼ 23 on 64 nodes to

---

[13] After recent successful streaming setups at 1024 nodes, the likely cause for this were scalability issues in the metadata strategy used in the openPMD-api.

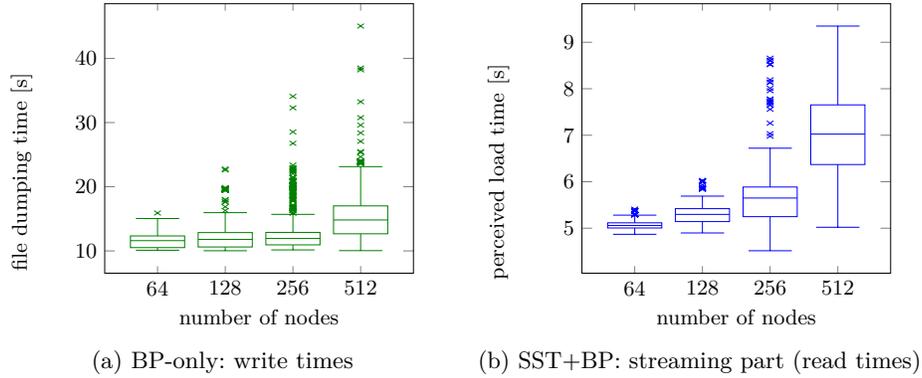

(a) BP-only: write times

(b) SST+BP: streaming part (read times)

Fig. 7: Perceived runtimes for file-based writes and streaming loads as boxplots. The box displays the interval containing 50% of sampled measurements. The upper whisker is at the largest value below "(upper quartile) $+ 1.5 \cdot$ IQR" (inter quartile difference), lower whisker accordingly.

$17 - 20$ on 512 nodes. In contrast, the SST+BP setup can increase the number of data dumps as long as the IO time can be hidden within the simulation time. Hence, we observe $32 \sim 34$ data dumps at 64 and at 128 nodes. Between 22 and 27 data dumps are written at 256 nodes and only $16 \sim 17$ at 512 nodes. This is because outputs are dropped as soon as the IO time cannot be hidden behind the simulation time any more.

We examine the portion of the simulation time that the IO plugin requires in PIConGPU (first percentage: raw IO operation, second percentage: IO plugin including host-side data preparation and reorganization). For the BP-only setup those numbers range from $(44\%/54\%)$ at 64 nodes to $(55\%/64\%)$ at 512 nodes, explaining the slight drop in successfully dumped simulation steps. For the streaming side of the SST+BP setup, they range from $(2.1\%/27\%)$ to $(6.2\%/32\%)$, showing that raw IO is barely noticeable at low scale, while gaining some significance due to communication latencies between up to 3072 writers in our setup.

As the BP engine in ADIOS2 has a feature to write data synchronously to node-local NVM drives and drain them asynchronously to the parallel filesystem, enabling this feature is possible for both setups benchmarked so far. However, we consistently measure worse throughputs achieved by doing so, most peculiarly a significant decrease in the performance of the SST engine. As noted in subsection 1.1, asynchronous draining can compete for network resources with other parts of a parallel application, i.e. inter-node communication. Such an effect on MPI and the SST engine is a possible reason for this observed behavior.

### 4.2   A staged simulation-analysis pipeline: Setup

The next benchmark loosely couples a PIConGPU simulation with an X-ray scattering analysis code named GAPD [5]. GAPD is an "atom-based polychromatic diffraction simulation code for direct, kinematics-based simulations of X-ray/ electron diffraction of large-scale atomic systems with mono-/polychromatic beams and arbitrary plane detector geometries" [5]. It takes into regard only particle data while ignoring mesh data produced by PIConGPU or similar simulations. Through scaling up to GPU clusters via MPI, GAPD is able to simulate diffraction patterns of systems up to 5 billion atoms.

Running this analysis is not only interesting from the perspective of extracting meaningful and interpretable data from the massive amounts of raw data produced by the simulation, it is also valuable as a means of reducing the amount of data to be processed via IO systems by several orders of magnitude down from the number of raw macroparticles to the number of points in reciprocal space.

GAPD is coupled with PIConGPU and configured to calculate the SAXS (Small-angle X-ray scattering) pattern from the input stream with the kinematical method [5]. If needed, the X-ray energy/wavelength, detection geometry can be adjusted in the input file whose configuration for the following benchmarks is found in the supplementary material along with software versions and an example for a created scatter plot [20]. This way, we commit ourselves to a realistic problem statement to be solved by simulation with PIConGPU and GAPD (scaled weakly to analyze scaling behavior) and aim to utilize the compute resources as fully as possible, avoiding waiting times introduced either by IO or asynchrony.

For reducing *IO-introduced* waiting times, IO should perform as fast as possible. To this end, the influence of chunk distribution strategies is shown. For reducing *asynchrony-introduced* waiting times, the setup will not block the simulation by performing IO. Again, ADIOS2 is configured to drop steps if the analysis has not finished yet, thus letting the pacing of the analysis determine the frequency of output. We demonstrate that this frequency can be tweaked by shifting the share of compute resources between writer and reader.

Codes must be scheduled carefully to allow for localized communication patterns. The Summit compute system hosts six GPUs per node, and the surveyed setup shares them equally between simulation and analysis, running three instances of PIConGPU and three instances of GAPD on each node. Distribution algorithms may or may not take the topology of the setup into account and we will show the impact of either.

Since GAPD only reads particle data, field data needs not be sent and does not influence the IO system, reducing the IO size per process to $\sim 3.1$ GiB. The field data stays relevant for computing the next steps in the PIC simulation.

### 4.3  A staged simulation-analysis pipeline: Evaluation

We evaluate the following three distribution strategies, based on the algorithms discussed in section 3.2:

**(1) Distribution by hostname:** Communication happens exclusively within a node. Distribution within a node is done via the Binpacking approach.
**(2) Binpacking:** Only the Binpacking algorithm runs and topology is ignored.

**(3) Slicing the dataset into hyperslabs:** The datasets are sliced into equal-sized hyperslabs which are distributed over the reading instances. Since PIConGPU uses no load balancing and data distribution in the problem space hence correlates with data distribution across the hardware, this approach keeps some notion of locality and avoids fully interconnected communication meshes.

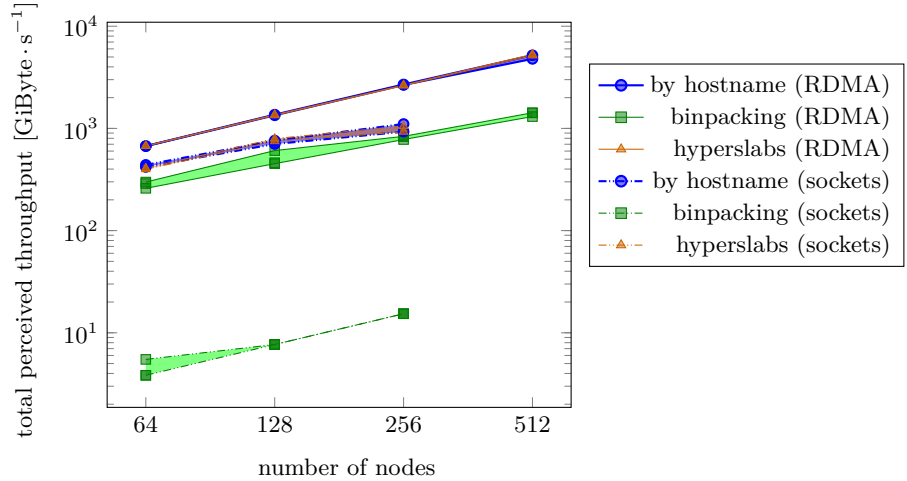

Fig. 8: Perceived total throughput. At 512 nodes, the RDMA-based methods reach 4.93, 1.35 and $5.12\,\text{TiByte} \cdot \text{s}^{-1}$ on average, respectively. The sockets-based methods reach 995, 15 and $985\,\text{GiByte} \cdot \text{s}^{-1}$. The *by hostname* and *hyperslabs* strategy results overlap each other, for RDMA and sockets alike.

Figure 8 shows the perceived throughput. As in section 4.1, the perceived throughput is defined by the time between write/load request and the end of the operation. This figure is subject to communication latencies, rendering the perceived throughput a lower bound for the actual throughput.

Again, each single benchmark is repeated three times and the plot shows the single measurements. The throughput is computed as the average over each parallel writer and each data exchange between writer and reader. The benchmark additionally shows the throughput observed with the sockets-based WAN implementation of the SST engine up until 256 nodes.

The RDMA tests show a reasonable quasi-linear parallel scaling behavior. The slightly higher peak bandwidth of $5.18\,\text{TiByte} \cdot \text{s}^{-1}$ compared to the bandwidths observed in subsection 4.1 can be related to the lower amount of data sent as well as to the different scheduling and communication patterns. Strategy *(2)* has a consistently worse performance than the other two strategies which clock out relatively similarly. The distinctive difference of strategy *(2)* is the number of communication partners that each parallel instance has, suggesting that controlling this number is important. Since no intra-host communication infrastructure is used, keeping communication strictly within one node in strategy *(1)* bears no measurable improvement over strategy *(3)*.

The WAN/sockets tests show similar tendencies, but at a consistently worse throughput than their RDMA-based counterparts. The worst throughput, with the Binpacking strategy, achieves throughputs between 400 and 970 $\text{GiByte} \cdot \text{s}^{-1}$, correlating to loading times up to and above three minutes. We conclude that for scientific simulations, where data is usually written in bulk, sockets do not provide a scalable streaming solution.

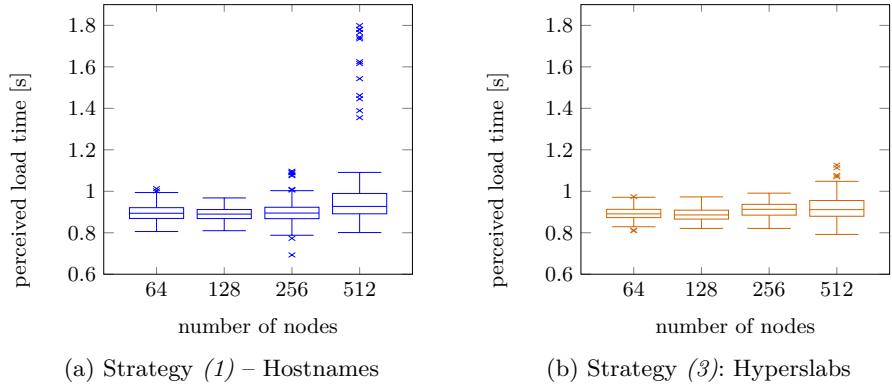

(a) Strategy *(1)* – Hostnames     (b) Strategy *(3)*: Hyperslabs

Fig. 9: Perceived data loading times for strategies *(1)* and *(3)* as boxplots.

Figure 9 plots the loading times for the two most promising strategies as boxplot. The median loading times are relatively consistent at 0.9 s for both strategies, explaining the good parallel scaling. The outliers observed for the Hostname strategy at 512 nodes all stem from the same single communication operation. Upon closer inspection, in that operation the Binpacking strategy (used within a node in strategy *(1)*) sent double the ideal amount of data to a single reader. This demonstrates that the worst-case behavior of the simple approximation algorithm can occur in practice. We conclude that for this setup, strategy *(3)* yields the best observed throughput and does not suffer from sporadic imbalances. Further heuristics should be explored for making the Binpacking algorithm more resistant against those.

With the setup chosen so far, GAPD takes around 5 minutes and 15 seconds to compute one scatter plot. Without blocking the simulation, this allows to create a scatter plot every 2000 steps in our experiments for the RDMA-based transport. Doing so via file-based loose coupling would require writing outputs of size 9.3 TiB per scatter plot at 512 nodes (after filtering the relevant data!). For the single data dump where the worst-case behavior of the Binpacking approach was observed, the respective scatter plot took roughly 10 minutes of computation due to the bad parallel load balancing.

For increasing the frequency of scatter plots, a benefit of loose coupling can be exploited: Dedicating five GPUs on a node to GAPD and only one to PIConGPU decreases the compute time for GAPD to roughly one minute, allowing it to run every 400 simulation steps. This is achieved only by changing the job script and without such a feature having been explicitly coded into either application, as would be necessary in a tightly-coupled monolithic application.

## 5   Summary & Outlook

The library openPMD-api has been extended to enable domain scientists a straight-forward transition from file-based to streaming IO. All data is scientifically self-describing and IO is performed flexibly, adapting to requirements from workflows and compute systems. Our benchmarks use streaming to exhaust available compute resources and to avoid bottlenecks caused by parallel filesystems. The achieved throughputs from the use of RDMA/Infiniband streaming reach double the bandwidth of the Summit PFS. In the first benchmark, this paper demonstrates the construction of a pipeline for asynchronous IO, and node-level data aggregation comes naturally with the setup. The second benchmark demonstrates the straightforward setup of a prototypical loosely-coupled simulation-analysis pipeline, avoiding to write intermediate results to persistent storage. The importance of data distribution is discussed and we respond to the challenge with flexibly interchangeable distribution algorithms, allowing to rapidly evaluate the best-performing strategy for a discrete setup. This approach allows for future extensibility towards setups with properties such as application-specific constraints or parallel load balancing that influence data distribution. Our experiments saw no benefits from using purely intra-node communication compared to inter-node communication. Backends with support for inter-process communication (IPC) techniques are a chance to exhaust potential from node-local communication, ideally to achieve near-perfect scaling at extreme scale. By substantiating that common technology like network sockets may not hold up to the challenge of streaming IO for HPC applications, this work employs efficient transport layers such as Infiniband successfully in an accessible manner for scientifically self-describing streaming data pipelines.

**Acknowledgements** This research used resources of the Oak Ridge Leadership Computing Facility at the Oak Ridge National Laboratory, which is supported by the Office of Science of the U.S. Department of Energy under Contract No. DE-AC05-00OR22725. Supported by the Exascale Computing Project (17-SC-20-SC), a collaborative effort of two U.S. Department of Energy organizations (Office of Science and the National Nuclear Security Administration). Supported by EC through Laserlab-Europe, H2020 EC-GA 871124. Supported by the Consortium for Advanced Modeling of Particles Accelerators (CAMPA), funded by the U.S. DOE Office of Science under Contract No. DE-AC02-05CH11231. This work was partially funded by the Center of Advanced Systems Understanding (CASUS), which is financed by Germany's Federal Ministry of Education and Research (BMBF) and by the Saxon Ministry for Science, Culture and Tourism (SMWK) with tax funds on the basis of the budget approved by the Saxon State Parliament.

From file-based HPC workflows to streaming data pipelines